\documentstyle{article}             

\catcode `\@=11
\def\numberbysection{\@addtoreset{equation}{section}
         \renewcommand{\theequation}{\thesection.\arabic{equation}}}
\numberbysection
\begin{document}
\begin{titlepage}
\def\thefootnote{\fnsymbol{footnote}}

\begin{flushleft}
\setlength{\baselineskip}{13pt}
ITP-UH-04/96 \hfill March 1996
\\
cond-mat/9604082
\end{flushleft}

\vspace*{\fill}
\begin{center}
{\large Algebraic Bethe Ansatz for $gl(2,1)$ Invariant 36--Vertex Models}

\vfill
\vspace{1.5 em}
{\sc Markus P.\ Pfannm\"uller}\footnote
               {e-mail: {\tt pfannm@itp.uni-hannover.de}}
and 
{\sc Holger Frahm}\footnote{e-mail: {\tt frahm@itp.uni-hannover.de}}\\
{\sl Institut f\"ur Theoretische Physik, Universit\"at Hannover,\\
D-30167~Hannover, Germany}\\
\vfill
ABSTRACT
\end{center}
\begin{quote}
Four dimensional irreducible representations of the superalgebra $gl(2,1)$
carry a free parameter.
We compute the spectra of the corresponding transfer matrices by means 
of the nested algebraic Bethe ansatz together with a generalized 
fusion procedure.
\end{quote}
\vfill
PACS-numbers:	71.27.+a~\      
		75.10.Lp~\      
		05.70.Jk~\      

\vfill
\setcounter{footnote}{0}
\end{titlepage}
\section{Introduction}
Recently vertex models built from representations of the superalgebra
$gl(2,1)$ 
or $q$-deformations thereof have attracted increasing
attention \cite{bglz:95,maas:95,esko:92,foka:93}.
One reason is their relation to integrable models of interacting
electrons in one spatial dimension: for example, the supersymmetric
$t$--$J$ model \cite{lai:74,suth:75,schl:87}
is obtained in the hamiltonian limit of the transfer matrix
for the vertex model based on the three dimensional fundamental
representation $[{1\over2}]_+$ of $gl(2,1)$ \cite{esko:92,foka:93}.
A special feature of this
algebra is the existence of a family of four dimensional representations
parametrized by a parameter $b\ne\pm{1\over2}$. From the corresponding
vertex model a one-parametric integrable model of interacting electrons can
be derived \cite{bglz:95}. This system of electrons with correlated
hopping has been solved recently by means of the co-ordinate Bethe Ansatz 
\cite{karn:94,befr:95}.

The relation to integrable vertex models provides an embedding of these
models into the framework of the Quantum Inverse Scattering Method.
However, up to now a direct solution by means of the nested algebraic Bethe
Ansatz has been obtained \emph{only} for the supersymmetric $t$--$J$
model \cite{esko:92} \cite{foka:93}.
The fusion method used to solve vertex models corresponding to
higher-dimensional representations of ordinary Lie algebras such
as $sl_n$ is not
applicable due to the peculiarities of the representation theory of a
superalgebra. Only for the model with $b={3\over2}$ this method can be
applied. Knowing the eigenvalues of the transfer matrix for this case
Maassarani \cite{maas:95} has made a conjecture for the general case
which is in agreement
with the spectrum of the hamiltonian obtained from the co-ordinate Bethe
Ansatz.

In this paper we extend the approach used by Maassarani to vertex models
built from $R$-matrices defined on tensor products of two \emph{different}
four dimensional representations $[b_1,{1\over2}]\otimes[b_2,{1\over2}]$ to
compute the eigenvalues of the corresponding transfer matrix
$\tau^{b_1b_2}(\mu)$: In the following section we give a short overview
over the three and four dimensional representations of $gl(2,1)$ that are
used together with the $R$-matrices acting on tensor products of these.
In Section 3 and 4
we obtain two inequivalent sets of Bethe Ansatz equations for this model
corresponding to a different choice of the reference state. Then we use the
fusion procedure to find the eigenvalues of the transfer matrix for
$b_1={3\over2}$. Then, using the set of Yang--Baxter equations for the
interwtiners between the different representations in addition with the
known analytic properties of the eigenvalues of the transfer matrix we can
determine the eigenvalues of $\tau^{b_1b_2}(\mu)$ up to an overall factor
which is fixed by studying the model built from two vertices only.
%
\section{{\boldmath$R$-matrices for $[\frac{1}{2}]_+$ 
          and $[b,\frac{1}{2}]$} representations}
In this section we present the $R$-matrices acting on tensorproducts 
of three dimensional $[\frac{1}{2}]_+$ and four dimensional $[b,\frac{1}{2}]$
representations of $gl(2,1)$ along with the corresponding Yang-Baxter
equations.
Before we discuss the particular form of the representations we introduce the
notation $[x]$ for the grading of an object $x$:
\begin{equation}
[x] = \left\{\begin{array}{cl}
		0 & \mbox{ if $x$ is bosonic (even)}\\
                1 & \mbox{ if $x$ is fermionic (odd)} 	     
	     \end{array}\right.
\end{equation}
The multiplication rule in graded tensor products differs from the ordinary
one by the appearance of additional signs.
For homogeneous elements $B$ and $v$ we get
\begin{equation}
(A \otimes B) (v \otimes w) = (-1)^{[B][v]}(Av)\otimes (Bw)
\end{equation}
Using homogeneous bases in the two vectorspaces this equation can
be written in components:
\begin{equation}
(A \otimes B)^{i_1,j_1}_{i_2,j_2} = (-1)^{([i_2]+[j_2])[j_1]}
A^{i_1,j_1}B^{i_2,j_2}
\end{equation}
The even part of the superalgebra $gl(2,1)$ consist of a the direct
sum of a $su(2)$ and a $u(1)$ Lie algebra.
Thus the basis vectors of the irreducible
representation can be labeled by the eigenvalue $B$ of the $u(1)$ operator,
the total spin and the z-component of the spin: $|B,S,S_z\rangle$
\cite{snr:77, marc:80I}.

The three dimensional representation $[\frac{1}{2}]_{+}$ contains a doublet
with $B=\frac{1}{2}$ and a singlet with $B=1$. 
We arrange the basis in the following order:
\begin{equation}
|{\scriptstyle\frac{1}{2}},\, {\scriptstyle\frac{1}{2}},\,
 {\scriptstyle\frac{1}{2}}\,\rangle ,\hspace{5mm}
|{\scriptstyle\frac{1}{2}},\, {\scriptstyle\frac{1}{2}}\,\,
 -{\scriptstyle\frac{1}{2}}\,\rangle ,\hspace{5mm}
|1, 0, 0\rangle 
\label{bas3}
\end{equation}
The first two states are considered as bosonic (grading 0), the last one
as fermionic (grading 1). For the supersymmetric $t$-$J$ model they
are identified with the electronic states with a spin up or a spin
down electron and an empty site. In this case one should choose the
opposite grading [1,1,0].

In the four dimensional $[b,\frac{1}{2}]$ representation we find a doublet
with $B=b$ and two singlets with $B=b\pm\frac{1}{2}$, respectively.
The basis is ordered according to
\begin{equation}
|b,\, {\scriptstyle\frac{1}{2}}\,\, {\scriptstyle\frac{1}{2}}\,\rangle ,
\hspace{3mm}
|b,\, {\scriptstyle\frac{1}{2}},\, -{\scriptstyle\frac{1}{2}}\,\rangle ,
\hspace{3mm}
|b-{\scriptstyle\frac{1}{2}},\, 0,\, 0\,\rangle ,\hspace{3mm}
|b+{\scriptstyle\frac{1}{2}},\, 0,\, 0\,\rangle 
\label{bas4}
\end{equation} 
Here the grading of the basis vectors is $[1,1,0,0]$. For the model with
correlated hopping they correspond to states with a single spin up or spin
down electron, an empty site and a doubly occupied site. The bosonic ones 
may be exchanged leading to an equivalent model. 

On the tensorproduct of two $[\frac{1}{2}]_{+}$ representations we have an $R$-matrix
$r^{33}$:
\begin{equation}
r^{33}(\lambda) = a(\lambda)\mbox{id}_9+ b(\lambda)\Pi_{33} 
\end{equation} 
Here $\mbox{id}_9$ denotes the $9 \times 9$ identity matrix and $\Pi_{33}$
is the graded permutation operator with matrix elements
$ (\Pi_{33})^{i_1,j_1}_{i_2,j_2} = (-1)^{[i_1][i_2]} \delta_{i_1,j_2}
\delta_{i_2,j_1}$. The functions $a$ and $b$ are given by
\begin{equation}
a(\lambda)=\frac{\lambda}{\lambda+1}, \hspace{2cm}
b(\lambda)=\frac{1}{\lambda+1}
\end{equation}
This $R$-matrix is a solution of the Yang-Baxter equation
\begin{equation}
r^{33}_{12}(\lambda-\mu)r^{33}_{13}(\lambda)r^{33}_{23}(\mu)  = 
r^{33}_{23}(\mu)r^{33}_{13}(\lambda)r^{33}_{12}(\lambda-\mu)
\label{YBeqr33r33}
\end{equation}
Here the lower indices denote the spaces in which the $R$-matrix acts.

The $R$-matrix $R^{3b}$ on the tensorproduct $[\frac{1}{2}]_+\otimes[b,\frac{1}{2}]$
can be constructed from the elementary
intertwiners of the irreducible components of the tensorproduct
\cite{dglz:95}. The result is:
\begin{equation}
R^{3b}(\mu) \sim It_1+\frac{2\mu-2b-1}{2\mu+2}It_2
\label{R3b}
\end{equation}
where $It_1$ and $It_2$ are the operators intertwining the eight dimensional
$[b+\frac{1}{2},1]$ subrepresentation and the four
dimensional $[b+1,\frac{1}{2}]$ 
subrepresentation, respectively. 

The tensorproduct $[b_1,\frac{1}{2}]\otimes[b_2,\frac{1}{2}]$ contains three
irreducible components, namely $[b_1+b_2,1]$ (D=8),
$[b_1+b_2+1,\frac{1}{2}]$ (D=4) and $[b_1+b_2-1,\frac{1}{2}]$ (D=4).
The R-matrix is given by the following combination of intertwiners:
\begin{equation}
R^{b_1b_2}(\mu) \sim It_1
                + \frac{4\mu-b_1-b_2-1}{4\mu+b_1+b_2+1}It_2 
                + \frac{4\mu+b_1+b_2-1}{4\mu-b_1-b_2+1}It_3
\label{R44}
\end{equation}
where $It_1$, $It_2$ and $It_3$ are the intertwiners for the eight and the
two four dimensional irreducible components, respectively.  

For the $R$-matrices defined in Eqs.\ (\ref{R3b}) and (\ref{R44})
the following Yang-Baxter equations hold:
\begin{eqnarray}
r^{33}_{12}(\lambda-\mu)R^{3b}_{13}(\lambda)R^{3b}_{23}(\mu) & = &
R^{3b}_{23}(\mu)R^{3b}_{13}(\lambda)r^{33}_{12}(\lambda-\mu)
\label{YBeqr33R34}\\
R^{3b_1}_{12}(\lambda-\mu)R^{3b_2}_{13}(\lambda)R^{b_1b_2}_{23}(\mu) & = &
R^{b_1b_2}_{23}(\mu)R^{3b2}_{13}(\lambda)R^{3b_1}_{12}(\lambda-\mu)
\label{YBeqR34R44}\\
R^{b_1b_2}_{12}(\lambda-\mu)R^{b_1b_3}_{13}(\lambda)R^{b_2b_3}_{23}(\mu) & = &
R^{b_2b_3}_{23}(\mu)R^{b_1b_3}_{13}(\lambda)R^{b_1b_2}_{12}(\lambda-\mu)
\label{YBeqR44R44}
\end{eqnarray}
Writing the Yang-Baxter eqautions in components one has to include additional
signs due to the grading:
\begin{equation}
r_{12}(\lambda-\mu)^{i_1,j_1}_{i_2,j_2}
R_{13}(\lambda)^{j_1,k_1}_{i_3,j_3}
R_{23}(\mu)^{j_2,k_2}_{j_3,k_3}
(-1)^{[j_2]([j_1]+[k_1])}
=
R_{23}(\mu)^{i_2,j_2}_{i_3,j_3}
R_{13}(\lambda)^{i_1,j_1}_{j_3,k_3}
r_{12}(\lambda-\mu)^{j_1,k_1}_{j_2,k_2}
(-1)^{[j_2]([i_1]+[j_1])}
\end{equation}
%
%
From the $R$-matrices for the different representations we can
construct monodromy matrices by taking matrix products in one component
of the tensorproduct -- the  auxillary or matrix space. 
\begin{equation}
T(\mu)^{a,\alpha_1,\ldots,\alpha_L}_{b,\beta_1,\ldots,\beta_L}
= R(\mu)^{a,a_L}_{\alpha_L,\beta_L}
  R(\mu)^{a_{L},a_{L-1}}_{\alpha_{L-1},\beta_{L-1}}\ldots
  R(\mu)^{a_3,a_2}_{\alpha_2,\beta_2}
  R(\mu)^{a_{2},a_{1}}_{\alpha_{1},\beta_{1}}
  (-1)^{\sum_{i=2}^{L}([\alpha_1]+[\beta_i])\sum_{j=1}^{i-1}[\alpha_i]}
\end{equation}
Again the grading gives rise to additional signs. As a consequence of
Eqs.\ (\ref{YBeqr33r33}) and (\ref{YBeqr33R34})-(\ref{YBeqR44R44})
the monodromy matrices
satisfy the following Yang-Baxter equations:
\begin{eqnarray}
r^{33}_{12}(\lambda-\mu)T^{33}_{13}(\lambda)T^{33}_{23}(\mu) & = &
T^{33}_{23}(\mu)T^{33}_{13}(\lambda)r^{33}_{12}(\lambda-\mu)
\label{YBeqt33t33}
\\
r^{33}_{12}(\lambda-\mu)T^{3b}_{13}(\lambda)T^{3b}_{23}(\mu) & = &
T^{3b}_{23}(\mu)T^{3b}_{13}(\lambda)r^{33}_{12}(\lambda-\mu)
\label{YBeqT34T34}
\\
R^{3b_1}_{12}(\lambda-\mu)T^{3b_2}_{13}(\lambda)T^{b_1b_2}_{23}(\mu) & = &
T^{b_1b_2}_{23}(\mu)T^{3b_2}_{13}(\lambda)R^{3b_1}_{12}(\lambda-\mu)
\label{YBeqT34T44}
\\
R^{b_1b_2}_{12}(\lambda-\mu)T^{b_1b_3}_{13}(\lambda)T^{b_2b_3}_{23}(\mu) & = &
T^{b_2b_3}_{23}(\mu)T^{b_1b_3}_{13}(\lambda)R^{b_1b_2}_{12}(\lambda-\mu)
\label{YBeqT44T44}
\end{eqnarray}
From the monodromy matrix the transfer matrix is obtained by taking
the supertrace in the auxillary space:
$\tau(\mu) = \sum_a (-1)^{[a]}T(\mu)^a_a$. 
The Yang-Baxter equations (\ref{YBeqt33t33}) and
(\ref{YBeqT34T34}-\ref{YBeqT44T44}) imply
that transfer matrices acting in the same quantum space commute:
\begin{eqnarray*}
\left[\tau^{33}(\lambda),\tau^{33}(\mu)\right]  & = & 0\\
\left[\tau^{3b}(\lambda),\tau^{3b}(\mu)\right] & =  & 0\\
\left[\tau^{3b_2}(\lambda),\tau^{b_1b_2}(\mu)\right] & =  & 0\\
\left[\tau^{b_1b_3}(\lambda),\tau^{b_2b_3}(\mu)\right] & = & 0
\end{eqnarray*}
These equations imply that $\tau^{3b}$, $\tau^{b_1b}$ and $\tau^{b_2b}$
share a system of common eigenvectors and thus can be diagonalized
simultaneously.
%
\section{Nested algebraic Bethe ansatz for {\boldmath$\tau^{3b}$}}
The transfer matrix $\tau^{3b}$ can be diagonalized directly by means of a
nested algebraic Bethe ansatz. The calculations can be performed in
close analogy to the NABA for the supersymmetric $t-J$-model \cite{esko:92}.
Thus we omit the details here and present only the most relevant steps.

We can represent $T^{3b}$ as a $3 \times 3$ matrix
in the auxillary space with entries beeing operators in the 
$N$-fold tensorproduct of four dimensional quantum spaces
\begin{equation}
T^{3b}(\lambda) = \left(\begin{array}{ccc}
                  A_{11}(\lambda) & A_{12 }(\lambda) & B_{1}(\lambda)\\
                  A_{21}(\lambda) & A_{22 }(\lambda) & B_{2}(\lambda)\\
                  C_{1}(\lambda) & C_{2 }(\lambda) & D(\lambda)\\
                  \end{array}\right)
\end{equation}
From the Yang-Baxter equation (\ref{YBeqT34T34}) we can derive commutation
relations for these quantum operators. The ones needed in the sequel are:
\begin{eqnarray}
D(\lambda)B_{i}(\mu) & = &
\frac{1}{a(\mu-\lambda)}B_{i}(\mu)D(\lambda)
+\frac{b(\lambda-\mu)}{a(\lambda-\mu)}B_{i}(\lambda)D(\mu)
\label{comDB}
\\
A_{i_2k_2}(\mu)B_{i_1}(\lambda) & = &
\frac{1}{a(\lambda-\mu)}r(\lambda-\mu)^{i_1,l_1}_{i_2,l_2}
B_{l_1}(\lambda)A_{l_2 k_2}(\mu)
-\frac{b(\lambda-\mu)}{a(\lambda-\mu)}B_{i_2}(\mu)A_{i_1k_2}(\lambda)
\label{comAB}
\\
B_{i_2}(\mu)B_{i_1}(\lambda) & = &
\frac{1}{b(\lambda-\mu)-a(\lambda-\mu)}r(\lambda-\mu)^{i_1,l_1}_{i_2,l_2}
B_{l_1}(\lambda)B_{l_2}(\mu)
\label{comBB}
\end{eqnarray}
Here $r(\mu)$ is the $R$-matrix of the rational six vertex model,
both states
being bosonic:
\begin{equation}
r(\mu) = a(\mu)\mbox{id}_4 + b(\mu)\Pi_{22}
\label{rnest}
\end{equation}
We choose the state
\begin{equation}
|0\rangle = \otimes^L |b-{\scriptstyle\frac{1}{2}},\,0,\,0\,\rangle
\label{refstat}
\end{equation} 
as reference state. The action of the monodromy matrix on this state is:
\begin{equation}
T^{3b}(\mu)|0\rangle = \left(\begin{array}{ccc}
			  1 & 0 & B_1(\mu)\\
                          0 & 1 & B_2(\mu)\\
                          0 & 0 & \left(\frac{4\mu-2b+5}{4\mu+2b+3}\right)^L
			  \end{array}\right) |0\rangle
\end{equation}
hence it is an eigenstate of $\tau^{3b}(\mu) =-D(\mu)+A_{11}(\mu)+A_{22}(\mu)$.
Starting from $|0\rangle$ we make the following ansatz for the Bethe vectors:
\begin{equation}
|\lambda_1,\ldots,\lambda_n\rangle = 
B_{a_1}(\lambda_1)\ldots B_{a_n}(\lambda_n)|0\rangle F^{a_n\ldots a_1}
\label{betvect}
\end{equation}
where summation over repeated indices is implied and the amplitudes
$F^{a_n\ldots a_1}$ are functions of the spectral parameters 
$\lambda_1,\ldots,\lambda_n$.
In order to calculate the action of the transfer matrix 
$\tau^{3b}(\mu)$ on such a state,
we use relations (\ref{comDB}) and
(\ref{comAB}) to commute the operators $D$ and $A$ through all $B$'s until
they hit the vacuum.
\begin{eqnarray}
D(\mu)|\lambda_1,\ldots,\lambda_n\rangle & = &
\left(\frac{4\mu-2b+5}{4\mu+2b+3}\right)^L 
\prod^{n}_{i=1}\frac{1}{a(\lambda_i-\mu)}
|\lambda_1,\ldots,\lambda_n\rangle 
\label{wtD}\\
& & +\sum_{k=1}^{n} (\tilde{\Lambda}_k)^{b_1,\ldots,b_n}_{a_1,\ldots,a_n}
F^{a_n,\ldots,a_1}B_{b_k}(\mu)
\prod_{j \neq k}^{n}B_{b_j}(\lambda_j)|0\rangle
\nonumber\\
{}[A_{11}(\mu)+A_{22}(\mu)]|\lambda_1,\ldots,\lambda_n\rangle & = &
\prod^{n}_{i=1}\frac{1}{a(\lambda_i-\mu)}
\tau^{(1)}(\mu)^{b_1,\ldots,b_n}_{a_1,\ldots,a_n}F^{a_n,\ldots,a_1}
\prod_{j=1}^{n}B_{b_j}(\lambda_j)|0\rangle
\label{wtA}
\\
& & +\sum_{k=1}^{n} (\Lambda_k)^{b_1,\ldots,b_n}_{a_1,\ldots,a_n}
F^{a_n,\ldots,a_1}B_{b_k}(\mu)
\prod_{j \neq k}^{n}B_{b_j}(\lambda_j)|0\rangle
\nonumber
\end{eqnarray}
Here $\tau^{(1)}$ is the transfer matrix of an inhomogeneous six vertex model
with $n$ sites that is constructed
from the $R$-matrix (\ref{rnest}).
\begin{equation}
\tau^{(1)}(\mu)^{b1,\ldots,b_n}_{a1,\ldots,a_n} = 
r(\lambda_n-\mu)^{d,d_n}_{b_n,a_n}
r(\lambda_{n-1}-\mu)^{d_n,d_{n-1}}_{b_{n-1},a_{n-1}}\ldots
r(\lambda_2-\mu)^{d_3,d_2}_{b_2,a_2}
r(\lambda_{1}-\mu)^{d_2,d}_{b_{1},a_{1}}
\end{equation}
The amplitudes $F^{a_n,\ldots,a_1}$ can now be identified with the components
of a vector $F$ in the state space of this $n$ site model.
As can be seen from Eqs.\ (\ref{wtD}) and (\ref{wtA}) a sufficient
condition for $|\lambda_1,\ldots,\lambda_n\rangle$
to be an eigenvector of $\tau(\mu)$ is that the unwanted terms
$\tilde{\Lambda}_k$ and $\Lambda_k$ cancel and that the vector $F$
is an eigenvector of the nested transfer matrix $\tau^{(1)}(\mu)$.

The condition that the unwanted terms $\tilde{\Lambda}_k$ and $\Lambda_k$
ought to cancel leads to a set of equations for the spectral parameters
$\lambda_k$: 
\begin{equation}
\left(\frac{4\lambda_k-2b+5}{4\lambda_k+2b+3}\right)^L F^{p_n,\ldots,p_1}
= \tau^{(1)}(\lambda_k)^{p_n,\ldots,p_1}_{a_n,\ldots,a_1} F^{a_n,\ldots,a_1}
\,,\hspace{1cm}k=1,\ldots,n
\label{BAeq1}
\end{equation}
In a second step (nesting) we have to diagonalize the transfer matrix 
$\tau^{(1)}$. This goal is achieved by another Bethe ansatz which gives the
well kown results for the inhomogeneous, rational six vertex model.
The amplitudes $F^{a_n,\ldots,a_1}$ are the components of the corresponding
eigenvectors. The eigenvalues are found to be:
\begin{equation}
\Lambda^{(1)}(\mu|\lambda_1,\ldots,\lambda_n|\nu_1,\ldots,\nu_{n_1}) = 
\prod_{i=1}^{n}a(\lambda_i-\mu)\prod_{j=1}^{n_1}\frac{1}{a(\nu_j-\mu)}
+\prod_{j=1}^{n_1}\frac{1}{a(\mu-\nu_j)}
\label{ewnest}
\end{equation}
The spectral parameters $\nu_j$ are subject to the following set of
Bethe equations:
\begin{equation}
\prod_{i=1}^{n}a(\lambda_i-\nu_j) = 
\prod_{k \neq j}^{n_1}\frac{a(\nu_k-\nu_j)}{a(\nu_j-\nu_k)},
\hspace{1cm} j=1,\ldots,n_1
\label{BAeqnest}
\end{equation}
%
If we insert the eigenvalues (\ref{ewnest}) of the nested transfer matrix
into Eq.\ (\ref{BAeq1}) we obtain the first set of Bethe equations:
\begin{equation}
\left(\frac{4\lambda_k-2b+5}{4\lambda_k+2b+3}\right)^L = 
\prod_{j=1}^{n_1}\frac{1}{a(\lambda_k-\nu_j)},\hspace{1cm} k=1,\ldots,n
\label{BAeqf1}
\end{equation}
From Eqs.\ (\ref{wtD}) and (\ref{wtA})
we can read of the eigenvalues of $\tau^{3b}(\mu)$
as now the eigenvalues of $\tau^{(1)}(\mu)$ are known:
\begin{eqnarray}
\Lambda^{3b}(\mu|\lambda_1,\ldots,\lambda_n|\nu_1,\ldots,\nu_{n_1}) & = & 
- \left(\frac{4\mu-2b+5}{4\mu+2b+3}\right)^L 
\prod_{i=1}^{n}\frac{1}{a(\lambda_i-\mu)}
\nonumber\\
& & +\prod_{j=1}^{n_1}\frac{1}{a(\nu_j-\mu)}
    +\prod_{j=1}^{n_1}\frac{1}{a(\mu-\nu_j)}
      \prod_{i=1}^{n}\frac{1}{a(\lambda_i-\mu)}
\label{EW1}
\end{eqnarray}
The eigenvalues and Bethe equations should be compared to the rational limit
of the corresponding equations in Ref.\ \cite{maas:95}. We find complete
agreement.
\section{A second Bethe ansatz}
%
In general the specific form of the eigenvalues and the Bethe equations
depends on the particular choice of the reference state from which the
Bethe vectors are built. For the transfer matrix $\tau^{3b}$ there exist
a second possibility besides (\ref{refstat}), namely the state:
\begin{equation}
|0\rangle=\otimes^{L}|b+{\scriptstyle\frac{1}{2}},\,0,\,0\,\rangle
\label{refstat2}
\end{equation}
The action of the monodromy matrix on this state is
\begin{equation}
T^{3b}(\mu)|0\rangle = \left(\begin{array}{ccc}
	\left(\frac{4\mu+2b-1}{4\mu+2b+3}\right)^L & 0 & 0\\
        0 & \left(\frac{4\mu+2b-1}{4\mu+2b+3}\right)^L & 0\\
        C_1(\mu) &C_2(\mu) & \left(\frac{4\mu-2b-3}{4\mu+2b+3}\right)^L
			  \end{array}\right) |0\rangle
\end{equation}
For the Bethe ansatz we need the following commutation relations between these
operators, which are derived from the Yang Baxter equation (\ref{YBeqT34T34}).
\begin{eqnarray}
D(\mu)C_{i}(\lambda) & = &
\frac{1}{a(\mu-\lambda)}C_{i}(\lambda)D(\mu)
+\frac{b(\lambda-\mu)}{a(\lambda-\mu)}C_{i}(\mu)D(\lambda)
\label{comDC}
\\
A_{i_1k_1}(\lambda)C_{k_2}(\mu) & = &
\frac{1}{a(\lambda-\mu)}r(\lambda-\mu)^{k_1,l_1}_{k_2,l_2}
C_{l_2}(\mu)A_{i_1 l_1}(\lambda)
-\frac{b(\lambda-\mu)}{a(\lambda-\mu)}C_{k_1}(\lambda)A_{i_1k_2}(\mu)
\label{comAC}
\\
C_{k_1}(\lambda)C_{k_2}(\mu) & = &
\frac{1}{b(\lambda-\mu)-a(\lambda-\mu)}r(\lambda-\mu)^{k_1,l_1}_{k_2,l_2}
C_{l_2}(\mu)B_{l_1}(\lambda)
\label{comCC}
\end{eqnarray}
Here $r(\mu)$ is again the $R$-matrix (\ref{rnest}) of the rational
six vertex state model, with two bosonic states.
Now we use the operators $C_i$ to to build the Bethe vectors starting
from the new reference state:
\begin{equation}
|\lambda_1,\ldots,\lambda_n\rangle = 
C_{a_1}(\lambda_1)\ldots C_{a_n}(\lambda_n)|0\rangle F^{a_n\ldots a_1}
\label{betvect2}
\end{equation}
The action of $\tau^{3b}$ on such a Bethe state is given by:
\begin{eqnarray}
D(\mu)|\lambda_1,\ldots,\lambda_n\rangle & = &
\left(\frac{4\mu-2b-3}{4\mu+2b+3}\right)^L 
\prod^{n}_{i=1}\frac{1}{a(\mu-\lambda_i)}
|\lambda_1,\ldots,\lambda_n\rangle 
\label{wtD2}\\
& & +\sum_{k=1}^{n} (\tilde{\Lambda}_k)^{b_1,\ldots,b_n}_{a_1,\ldots,a_n}
F^{a_n,\ldots,a_1}B_{b_k}(\mu)
\prod_{j \neq k}^{n}B_{b_j}(\lambda_j)|0\rangle
\nonumber\\
{}[A_{11}(\mu)+A_{22}(\mu)]|\lambda_1,\ldots,\lambda_n\rangle & = &
\left(\frac{4\mu+2b-1}{4\mu+2b+3}\right)^L 
\prod^{n}_{i=1}\frac{1}{a(\mu-\lambda_i)}
\tau^{(1)}(\mu)^{b_1,\ldots,b_n}_{a_1,\ldots,a_n}F^{a_n,\ldots,a_1}
\prod_{j=1}^{n}B_{b_j}(\lambda_j)|0\rangle
\label{wtA2}
\\
& & +\sum_{k=1}^{n} (\Lambda_k)^{b_1,\ldots,b_n}_{a_1,\ldots,a_n}
F^{a_n,\ldots,a_1}B_{b_k}(\mu)
\prod_{j \neq k}^{n}B_{b_j}(\lambda_j)|0\rangle
\nonumber
\end{eqnarray}
Here $\tau^{(1)}$ is the transfermatrix of an inhomogeneous $n$ site 
model that is constructed from the $R$-matrix (\ref{rnest}).
\begin{equation}
\tau^{(1)}(\mu)^{b1,\ldots,b_n}_{a1,\ldots,a_n} = 
r(\mu-\lambda_n)^{d,d_n}_{b_n,a_n}
r(\mu-\lambda_{n-1})^{d_n,d_{n-1}}_{b_{n-1},a_{n-1}}\ldots
r(\mu-\lambda_2)^{d_3,d_2}_{b_2,a_2}
r(\mu-\lambda_{1})^{d_2,d}_{b_{1},a_{1}}
\end{equation}
The condition that the unwanted terms ought to cancel leads to the following
set of equations:
\begin{equation}
\left(\frac{4\lambda_k-2b-3}{4\lambda_k+2b-1}\right)^L F^{p_n,\ldots,p_1}
= \tau^{(1)}(\lambda_k)^{p_n,\ldots,p_1}_{a_n,\ldots,a_1} F^{a_n,\ldots,a_1}
\,,\hspace{1cm}k=1,\ldots,n
\label{BAeq2}
\end{equation}
As before the nested transfer matrix $\tau^{(1)}$ is diagonalized by a
second Bethe ansatz.
The corresponding eigenvalues are found to be:
\begin{equation}
\Lambda^{(1)}(\mu|\lambda_1,\ldots,\lambda_n|\nu_1,\ldots,\nu_{n_1}) = 
\prod_{i=1}^{n}a(\mu-\lambda_i)\prod_{j=1}^{n_1}\frac{1}{a(\mu-\nu_j)}
+\prod_{j=1}^{n_1}\frac{1}{a(\nu_j-\mu)}
\label{ewnest2}
\end{equation}
where $\nu_j$ are solutions of the following set of Bethe equations
\begin{equation}
\prod_{i=1}^{n}a(\nu_j-\lambda_i) = 
\prod_{k \neq j}^{n_1}\frac{a(\nu_j-\nu_k)}{a(\nu_k-\nu_j)},
\hspace{1cm} j=1,\ldots,n_1
\label{BAeqnest2}
\end{equation}
If we insert the eigenvalues $\lambda^{(1)}$
into Eqs.\ (\ref{BAeq2}) we obtain the first level Bethe equations: 
\begin{equation}
\left(\frac{4\lambda_k-2b-3}{4\lambda_k+2b-1}\right)^L = 
\prod_{j=1}^{n_1}\frac{1}{a(\nu_j-\lambda_k)},\hspace{1cm} k=1,\ldots,n
\label{BAeqf2}
\end{equation}
From Eqs.\ (\ref{wtD2}) and (\ref{wtA2}) we can now determine the
eigenvalues of the transfer matrix $\tau^{3b}(\mu)$:
\begin{eqnarray}
\Lambda^{3b}(\mu|\lambda_1,\ldots,\lambda_n|\nu_1,\ldots,\nu_{n_1}) & = & 
- \left(\frac{4\mu-2b-3}{4\mu+2b+3}\right)^L 
\prod_{i=1}^{n}\frac{1}{a(\mu-\lambda_i)}
\nonumber\\
& &+ \left(\frac{4\mu+2b-1}{4\mu+2b+3}\right)^L
 \left(\prod_{j=1}^{n_1}\frac{1}{a(\mu-\nu_j)}
    +\prod_{j=1}^{n_1}\frac{1}{a(\nu_j-\mu)}
      \prod_{i=1}^{n}\frac{1}{a(\mu-\lambda_i)}\right)
\label{EW2}
\end{eqnarray}
This completes the the second Bethe ansatz. We postpone the discussion of
the relation between the two Bethe ans{\"a}tze to Section 6. 
\section{Fusion}
%
The tensor product of two $[\frac{1}{2}]_{+}$ representations contains
a four dimensional $[b=\frac{3}{2},\frac{1}{2}]$ and five dimensional 
$[1]_{+}$ representation.
We construct the fused $R$-matrix for the ``sacttering'' of two
composite ``$[\frac{1}{2}]_{+}$-particles'' with a third ``$[b,\frac{1}{2}]$
 particle'' from
\begin{equation}
R_{12,3} (\lambda) = R^{3b}_{13}(\lambda+\lambda_0)
                     R^{3b}_{23}(\lambda-\lambda_0)
\end{equation}
In general $[\frac{3}{2},\frac{1}{2}]$ and $[1]_{+}$ states
are mixed by the scattering.
However the $[\frac{3}{2},\frac{1}{2}]$ state will not be destroyed if the 
triangularity condition holds:
\begin{equation}
P_{12}^{5} R_{12,3}P_{12}^{\frac{3}{2}} = 0
\label{triang}
\end{equation}
Here $P_{12}^{5}$ and $P_{12}^{\frac{3}{2}}$ denote the projectors onto
the five dimensional representation $[1]_{+}$ and the four dimensional
$[\frac{3}{2},\frac{1}{2}]$ representation, respectively.
$r^{33}$ becomes proportional to a projector onto the
five dimensional representation for the special value 1 of the spectral
parameter
\begin{equation}
r^{33}_{12}(1) \sim P^{5}_{12}
\end{equation}
From the Yang-Baxter equation (\ref{YBeqr33R34}) we obtain 
\begin{equation}
P^{5}_{12}R^{34}_{13}(\lambda+1/2)R^{3b}_{23}(\lambda-1/2) =
R^{3b}_{2b}(\lambda-1/2)R^{3b}_{13}(\lambda+1/2)P^{5}_{12}
\label{YBeqr33R34spec}
\end{equation}
This allows to prove the triangularity condition (\ref{triang}) for $R_{12,3}$
for $\lambda_0=\frac{1}{2}$.
Thus the matrix
\begin{equation}
R_{(12),3} = P_{12}^{\frac{3}{2}}R_{12,3}P_{12}^{\frac{3}{2}}
\end{equation}
indeed describes the ``scattering'' of a $[\frac{3}{2},\frac{1}{2}]$ particle
with a $[b,\frac{1}{2}]$ particle. A Yang-Baxter equation holds for this
$R$-matrix:
\begin{equation}
R_{(12),3}(\lambda-\mu)R_{(12),4}(\lambda)R^{3b}_{34}(\mu) =
R^{3b}_{34}(\mu)R_{(12),4}(\lambda)R_{(12),3}(\lambda-\mu)
\end{equation}
The triangularity condition (\ref{triang}) states that
$R_{12,3}$ becomes an upper block-triangular matrix if we change the basis
in the tensor product $1 \otimes 2$ such that the first 4 vectors form
a basis for $[\frac{3}{2},\frac{1}{2}]$ and the next
five a basis for $[1]_{+}$. 
Let $B$ be the corresponding transformation matrix. Then we have:
\begin{equation}
B^{-1}R_{12,3}B = \left(\begin{array}{cc}
	   R^{\frac{3}{2}b} & *\\
	      0    & R^{5b}
	   \end{array}\right)
\label{R12_3}
\end{equation}
We can replicate this $R$-matrix to a chain of $L$ sites (the auxillary space
is the tensorproduct $1 \otimes 2$!). Taking the trace over spaces 1 and 2
leads to the following relation between the transfer matrices of the various 
models:
\begin{equation}
\tau^{3b}(\lambda+{\scriptstyle\frac{1}{2}})\,
\tau^{3b}(\lambda-{\scriptstyle\frac{1}{2}}) =
\tau^{\frac{3}{2}b}(\lambda)+\tau^{5b}(\lambda)
\label{fustau}
\end{equation}
Thus for the eigenvalues of the transfer matrices we find the relation
\begin{equation}
\Lambda^{3b}(\lambda+{\scriptstyle\frac{1}{2}})\,
\Lambda^{3b}(\lambda-{\scriptstyle\frac{1}{2}}) = 
\Lambda^{\frac{3}{2}b}(\lambda)+\Lambda^{5b}(\lambda)
\label{fusEW}
\end{equation}
Unfortunately there are two unknowns in this equation, namely
$\Lambda^{\frac{3}{2}b}(\lambda)$ and $\Lambda^{5b}(\lambda)$.
We expect the eigenvalues
corresponding to Bethe vectors (\ref{betvect}) 
to be a sum of the vacuum expectation
values of the diagonal operators of the transfer matrix dressed by some
factors that depend on the spectral parameters of the Bethe ansatz state.
However this is not sufficient to assign the various parts of the LHS of
(\ref{fusEW}) to $\Lambda^{\frac{3}{2}b}(\lambda)$ and $\Lambda^{5b}(\lambda)$
because the transfer matrices have common vacuum expectation values. 
As second argument we use the analyticity of the eigenvalues. The Bethe
ansatz equations (\ref{BAeqnest}) and (\ref{BAeqf1}) 
are exactly the conditions that
the eigenvalues are analytic functions of $\mu$. 
The diagonal parts of monodromy matrix $T^{\frac{3}{2}b}$ have
the following four eigenvalues on the reference state $|0\rangle$ 
(\ref{refstat}): 
\begin{eqnarray}
T^{\frac{3}{2}b}_{11}(\mu)|0\rangle =
T^{\frac{3}{2}b}_{22}(\mu)|0\rangle 
& = & \left(\frac{2\mu-b+\frac{3}{2}}{2\mu+b+\frac{1}{2}}\right)^L |0\rangle
\nonumber\\
T^{\frac{3}{2}b}_{33}(\mu)|0\rangle & = & 1|0\rangle
\label{T34}\\
T^{\frac{3}{2}b}_{44}(\mu)|0\rangle & = & 
\left(\frac{2\mu-b+\frac{7}{2}}{2\mu+b+\frac{5}{2}}\,
      \frac{2\mu-b+\frac{3}{2}}{2\mu+b+\frac{1}{2}}\right)^L|0\rangle
\nonumber
\end{eqnarray}
For $T^{54}$ we find
\begin{eqnarray}
T^{5b}_{11}(\mu)|0\rangle = 
T^{5b}_{22}(\mu)|0\rangle = 
T^{5b}_{33}(\mu)|0\rangle & = & 1|0\rangle
\nonumber\\
T^{5b}_{44}(\mu)|0\rangle  = 
T^{5b}_{55}(\mu)|0\rangle & = & 
 \left(\frac{2\mu-b+\frac{7}{2}}{2\mu+b+\frac{5}{2}}\right)^L |0\rangle
\label{T54}
\end{eqnarray}
Inserting the eigenvalues (\ref{EW1}) into Eq.\ (\ref{fusEW})  we obtain:
\begin{eqnarray}
\Lambda^{3b}\left(\mu-\frac{1}{2}\right)
\Lambda^{3b}\left(\mu+\frac{1}{2}\right)& =  &
- \left(\frac{2\mu-b+\frac{3}{2}}{2\mu+b+\frac{1}{2}}\right)^L 
\Big\{
 \prod_{i=1}^{n}\frac{\mu-\lambda_i-\frac{3}{2}}{\mu-\lambda_i-\frac{1}{2}}
 \prod_{j=1}^{n_1}\frac{\mu-\nu_j-\frac{1}{2}}{\mu-\nu_j+\frac{1}{2}}
\nonumber\\& & \hspace{3cm}
+ \prod_{i=1}^{n}\frac{\mu-\lambda_i-\frac{3}{2}}{\mu-\lambda_i+\frac{1}{2}}
 \prod_{j=1}^{n_1}\frac{\mu-\nu_j+\frac{3}{2}}{\mu-\nu_j+\frac{1}{2}}
\Big\}
\nonumber\\
& & + \left(\frac{2\mu-b+\frac{3}{2}}{2\mu+b+\frac{1}{2}}\,
     \frac{2\mu-b+\frac{7}{2}}{2\mu+b+\frac{5}{2}}\right)^L 
\prod_{i=1}^{n}\frac{\mu-\lambda_i-\frac{3}{2}}{\mu-\lambda_i+\frac{1}{2}}
\nonumber\\
& & - \left(\frac{2\mu-b+\frac{7}{2}}{2\mu+b+\frac{5}{2}}\right)^L 
\Big\{
 \prod_{i=1}^{n}\frac{\mu-\lambda_i-\frac{1}{2}}{\mu-\lambda_i+\frac{1}{2}}
 \prod_{j=1}^{n_1}\frac{\mu-\nu_j-\frac{3}{2}}{\mu-\nu_j-\frac{1}{2}}
\nonumber\\ & &\hspace{3cm}
 + \prod_{i=1}^{n}\frac{\mu-\lambda_i-\frac{3}{2}}{\mu-\lambda_i+\frac{1}{2}}
  \prod_{j=1}^{n_1}\frac{\mu-\nu_j+\frac{1}{2}}{\mu-\nu_j-\frac{1}{2}}
\Big\}
\nonumber\\
& & + 1 \, \Big\{
 \prod_{i=1}^{n}\frac{\mu-\lambda_i-\frac{3}{2}}{\mu-\lambda_i-\frac{1}{2}}
+ \prod_{j=1}^{n_1}\frac{\mu-\nu_j-\frac{3}{2}}{\mu-\nu_j+\frac{1}{2}}
+ \prod_{i=1}^{n}\frac{\mu-\lambda_i-\frac{3}{2}}{\mu-\lambda_i+\frac{1}{2}}
 \prod_{j=1}^{n_1}\frac{\mu-\nu_j+\frac{3}{2}}{\mu-\nu_j-\frac{1}{2}}
\nonumber\\ 
& & \hspace{5mm}+ \prod_{i=1}^{n}\frac{\mu-\lambda_i-\frac{1}{2}}
		   {\mu-\lambda_i+\frac{1}{2}}
 \prod_{j=1}^{n_1}\frac{\mu-\nu_j-\frac{3}{2}}{\mu-\nu_j-\frac{1}{2}}
	   \frac{\mu-\nu_j+\frac{3}{2}}{\mu-\nu_j+\frac{1}{2}}
\Big\}
\end{eqnarray}
The first two terms clearly belong to $\Lambda^{\frac{3}{2}b}(\mu)$ and
the third one to  $\lambda^{5b}(\mu)$. Checking the residues at the poles
$\mu= \lambda_i+\frac{1}{2}$, $\mu= \lambda_i-\frac{1}{2}$ and
$\mu= \nu_j+\frac{1}{2}$ it can be easily seen that 
$\prod_{i=1}^{n}\frac{\mu-\lambda_i-\frac{3}{2}}{\mu-\lambda_i-\frac{1}{2}}$
is the missing part of $\Lambda^{\frac{3}{2}b}$. Thus we have 
\begin{eqnarray}
\Lambda^{\frac{3}{2}b}(\mu) & = &
- \left(\frac{2\mu-b+\frac{3}{2}}{2\mu+b+\frac{1}{2}}\right)^L\, 
\Big\{
 \prod_{i=1}^{n}\frac{\mu-\lambda_i-\frac{3}{2}}{\mu-\lambda_i-\frac{1}{2}}
 \prod_{j=1}^{n_1}\frac{\mu-\nu_j-\frac{1}{2}}{\mu-\nu_j+\frac{1}{2}}
+ \prod_{i=1}^{n}\frac{\mu-\lambda_i-\frac{3}{2}}{\mu-\lambda_i+\frac{1}{2}}
 \prod_{j=1}^{n_1}\frac{\mu-\nu_j+\frac{3}{2}}{\mu-\nu_j+\frac{1}{2}}
\Big\}
\nonumber\\
 & &  + \left(\frac{2\mu-b+\frac{3}{2}}{2\mu+b+\frac{1}{2}}
    \frac{2\mu-b+\frac{7}{2}}{2\mu+b+\frac{5}{2}}\right)^L \,
\prod_{i=1}^{n}\frac{\mu-\lambda_i-\frac{3}{2}}{\mu-\lambda_i+\frac{1}{2}}
\,\,
 +  \prod_{i=1}^{n}\frac{\mu-\lambda_i-\frac{3}{2}}{\mu-\lambda_i-\frac{1}{2}}
\label{EW3_2b}
\end{eqnarray}
%
%
The situation encountered so far is not yet satisfactory as one wishes to
diagonalize the transfer matrix for a $[b_1,\frac{1}{2}]$ representation in the
auxiallary space and a $[b_2,\frac{1}{2}]$ representation in the quantum space.
With the current results we are limited to the case $b_1=\frac{3}{2}$.
Especially we cannot handle the situation of
$[b,\frac{1}{2}] \otimes [b,\frac{1}{2}]$,
which gives rise to models of correlated electrons with an additional
free parameter.

From the fact that $\tau^{\frac{3}{2}b_2}(\mu)$ and
$\tau^{b_1b_2}(\lambda)$ commute and thus share a system of common
eigenvectors, we conclude that the Bethe ansatz equations 
(\ref{BAeqnest}) and (\ref{BAeqf1}) must be preserved.

We modify the eigenvalues (\ref{EW3_2b}) by replacing the 
vacuum expectation values of the diagonal elments of $T^{\frac{3}{2}b_2}$
(\ref{T34}) by those of $T^{b_1b_2}$ and then modify
the ``dressing factors'' such
that the Bethe ansatz equations still ensure the analyticity.

Examining the second term of (\ref{EW3_2b}), we see that the singularities 
of the ``dressing factor'' must be at values of $\mu$ such that
$\frac{2\mu+b_1-b_2+2}{2\mu+b_1+b_2+1}=
\frac{2\lambda_k-b_1+\frac{5}{2}}{2\lambda_k+b_1+\frac{3}{2}}$
in order to lead to the Bethe ansatz equation (\ref{BAeqf1}). This 
fixes the denominator of the ``dressing factor'' to be 
$2\mu-2\lambda_i+b_1-\frac{1}{2}$.
A similar factor comes with the vacuum expectation
value $\frac{2\mu+b_1-b_2+2}{2\mu+b_1+b_2+1}$. The right hand side of
(\ref{BAeqf1}) then is used to fix the ``nested dressing factor''. 
A similar argument works for the last term, where the singulatity has to be
such that $\frac{2\mu+b_1-b_2}{2\mu+b_1+b_2-1}=
\frac{2\lambda_k-b_1+\frac{5}{2}}{2\lambda_k+b_1+\frac{3}{2}}$. 

Only the numerator of the product $\prod_{i=1}^n$ still remains unkown. 
We use the first non trivial solution of the BAeq's for
a two site model, namely $n=1,n_1=0,\lambda_1=-1$ and determine the
corresponding eigenvalue by operating with the transfer matrix on the
corresponding Bethe vector. Evaluating the result at $\mu=(b2-b1)/2$
isolates the last term. This leads to the final result for 
the eigenvalues of $\tau^{b_1b_2}$:
\begin{eqnarray}
\Lambda^{b_1b_2}(\mu) & = &
- \left(\frac{2\mu+b_1-b_2}{2\mu+b_1+b_2-1}\right)^L 
\Big\{
 \prod_{i=1}^{n}\frac{2\mu-2\lambda_i-b_1-\frac{3}{2}}
	      {2\mu-2\lambda_i+b_1-\frac{5}{2}}
 \prod_{j=1}^{n_1}\frac{2\mu-2\nu_j+b_1-\frac{5}{2}}
		{2\mu-2\nu_j+b_1-\frac{1}{2}}
\nonumber\\
& & + \prod_{i=1}^{n}\frac{2\mu-2\lambda_i-b_1-\frac{3}{2}}
		   {2\mu-2\lambda_i+b_1-\frac{1}{2}}
 \prod_{j=1}^{n_1}\frac{2\mu-2\nu_j+b_1+\frac{3}{2}}
		{2\mu-2\nu_j+b_1-\frac{1}{2}}
\Big\}
\nonumber\\
& &  + \left(\frac{2\mu+b_1-b_2}{2\mu+b_1+b_2-1}\,
    \frac{2\mu+b_1-b_2+2}{2\mu+b_1+b_2+1}\right)^L\, 
\prod_{i=1}^{n}\frac{2\mu-2\lambda_i-b_1-\frac{3}{2}}
	     {2\mu-2\lambda_i+b_1-\frac{1}{2}}
\nonumber\\
& & +  \prod_{i=1}^{n}\frac{2\mu-2\lambda_i-b_1-\frac{3}{2}}
		   {2\mu-2\lambda_i+b_1-\frac{5}{2}}
\label{EWb1b2}
\end{eqnarray}
We checked the result numerically. For $b_1=b_2$ the equations are 
equivalent to the rational limit of the conjectures in
Ref.\ \cite{maas:95}. Taking the logarithmic derivative of the eigenvalues
we find the energies that where calculated in Ref.\ \cite{befr:95} by
means of co-ordinate Bethe ansatz. 

We can proceed with results from the second Bethe ansatz in a similar
manner. To keep the presentation short we will only give the
corresponding results for the more general case of inhomogeneous chains 
in Sec.\ 6. 
\section{Inhomogeneous chains}
%
A straight forward generalisation is now the construct a monodromy matrix
from $R^{bb_i}$ matrices, i.e. we choose the
representation $[b,\frac{1}{2}]$ in
the auxillary space and the representation $[b_i,\frac{1}{2}]$ in the quantum
space at site $i$. This model is also integrable by construction because of
the Yang-Baxter equation (\ref{YBeqR44R44}).
Modifying the vacuum expectation
values, the eigenvalues and Bethe ansatz equations can be derived from
the ones obtained in the previous section.
It is convenient to rescale and
shift the spectral parameters according to
$\lambda_k \rightarrow -i\lambda_k-1$,
$\nu_j \rightarrow -i\nu_j+\frac{1}{2}$:
\begin{eqnarray}
\Lambda^{b\{b_k\}}(\mu) & = &
- \prod_{k=1}^{L}\frac{2\mu+b-b_k}{2\mu+b+b_k-1}  
\Big\{
 \prod_{i=1}^{n}\frac{2\mu+2i\lambda_i-b+\frac{1}{2}}
	      {2\mu+2i\lambda_i+b-\frac{1}{2}}
 \prod_{j=1}^{n_1}\frac{2\mu+2i\nu_j+b-\frac{3}{2}}
		{2\mu+2i\nu_j+b+\frac{1}{2}}
\nonumber\\
& & + \prod_{i=1}^{n}\frac{2\mu+2i\lambda_i-b+\frac{1}{2}}
		   {2\mu+2i\lambda_i+b+\frac{3}{2}}
 \prod_{j=1}^{n_1}\frac{2\mu+2i\nu_j+b+\frac{5}{2}}
		{2\mu+2i\nu_j+b+\frac{1}{2}}
\Big\}
\nonumber\\
& & + \prod_{k=1}^{L}\left(\frac{2\mu+b-b_k}{2\mu+b+b_k-1}\,
		    \frac{2\mu+b-b_k+2}{2\mu+b+b_k+1}\right) 
\prod_{i=1}^{n}\frac{2\mu+2i\lambda_i-b+\frac{1}{2}}
	     {2\mu+2i\lambda_i+b+\frac{3}{2}}
\nonumber\\
& & +  \prod_{i=1}^{n}\frac{2\mu+2i\lambda_i-b+\frac{1}{2}}
		   {2\mu+2i\lambda_i+b-\frac{1}{2}}
\label{EWimp}
\end{eqnarray}
The Bethe equations are
\begin{equation}
\prod_{k=1}^{L}\frac{\lambda_l-i(\frac{b_k}{2}-\frac{1}{4})}
	     {\lambda_l+i(\frac{b_k}{2}-\frac{1}{4})} = 
\prod_{j=1}^{n_1}\frac{\lambda_l-\nu_j+\frac{i}{2}}
	       {\lambda_l-\nu_j-\frac{i}{2}},\hspace{1cm} l=1,\ldots,n
\label{BAeqimp1}
\end{equation}
\begin{equation}
\prod_{l=1}^{n}\frac{\lambda_l-\nu_i+\frac{i}{2}}
	     {\lambda_l-\nu_i-\frac{i}{2}} = 
- \prod_{j=1}^{n_1}\frac{\nu_j-\nu_i+i}{\nu_i-\nu_j-i},
\hspace{1cm} i=1,\ldots,n_1
\label{BAeqimpnest}
\end{equation}
In a similar fashion we can proceed with the results from the second Bethe
ansatz. Finally we obtain:
\begin{eqnarray}
\Lambda^{b\{b_k\}}(\mu) & = &
- \prod_{k=1}^{L}\left(\frac{2\mu-b+b_k}{2\mu+b+b_k-1}  
		\frac{2\mu-b-b_k+1}{2\mu+b+b_k+1}\right) 
\Big\{
 \prod_{i=1}^{n}\frac{2\mu+2i\lambda_i+b+\frac{1}{2}}
	      {2\mu+2i\lambda_i-b-\frac{1}{2}}
 \prod_{j=1}^{n_1}\frac{2\mu+2i\nu_j-b-\frac{3}{2}}
		{2\mu+2i\nu_j-b+\frac{1}{2}}
\nonumber\\
& & + \prod_{i=1}^{n}\frac{2\mu+2i\lambda_i+b+\frac{1}{2}}
		   {2\mu+2i\lambda_i-b+\frac{3}{2}}
 \prod_{j=1}^{n_1}\frac{2\mu+2i\nu_j-b+\frac{5}{2}}
		{2\mu+2i\nu_j-b+\frac{1}{2}}
\Big\}
\nonumber\\
& & + \prod_{k=1}^{L}\left(\frac{2\mu-b+b_k}{2\mu+b+b_k-1}\,
		    \frac{2\mu-b+b_k+2}{2\mu+b+b_k+1}\right) 
\prod_{i=1}^{n}\frac{2\mu+2i\lambda_i+b+\frac{1}{2}}
	     {2\mu+2i\lambda_i-b+\frac{3}{2}}
\nonumber\\
& & + \prod_{k=1}^{L}\left(\frac{2\mu-b-b_k+1}{2\mu+b+b_k-1}\,
		    \frac{2\mu-b-b_k-1}{2\mu+b+b_k+1}\right) 
      \prod_{i=1}^{n}\frac{2\mu+2i\lambda_i+b+\frac{1}{2}}
		   {2\mu+2i\lambda_i-b-\frac{1}{2}}
\label{EWimp2}
\end{eqnarray}
The Bethe equations are
\begin{equation}
\prod_{k=1}^{L}\frac{\lambda_l-i(\frac{b_k}{2}+\frac{1}{4})}
	     {\lambda_l+i(\frac{b_k}{2}+\frac{1}{4})} = 
\prod_{j=1}^{n_1}\frac{\lambda_l-\nu_j+\frac{i}{2}}
	       {\lambda_l-\nu_j-\frac{i}{2}},\hspace{1cm} l=1,\ldots,n
\label{BAeqimp12}
\end{equation}
\begin{equation}
\prod_{l=1}^{n}\frac{\lambda_l-\nu_i+\frac{i}{2}}
	     {\lambda_l-\nu_i-\frac{i}{2}} = 
- \prod_{j=1}^{n_1}\frac{\nu_j-\nu_i+i}{\nu_i-\nu_j-i},
\hspace{1cm} i=1,\ldots,n_1
\label{BAeqimpnest2}
\end{equation}
This second set of Bethe equations can be obtained from the first set
(\ref{BAeqimp1}-\ref{BAeqimpnest}) by replacing $b_k$ by $-b_k$.
If in addition $b$ is replaced by $-b$ the eigenvalues (\ref{EWimp})
and (\ref{EWimp2}) are found to be equal up to an overall factor.
This behaviour is explained by the fact that there exist an automorphism
of the superalgebra $gl(2,1)$ which maps the $u(1)$ operator $B$ onto
$-B$.
\section{Discussion}
%
%
In this paper we have computed the spectrum of vertex models invariant
under the action of the superalgebra $gl(2,1)$ by means of the Bethe
Ansatz. Depending on the choice of the reference state in the four
dimensional quantum space of the local vertices two different Bethe
Ans\"atze are possible which are related to the automorphism of the
superalgebra which maps the corresponding states onto each other. The
solutions start from one of the \emph{bosonic} highest weigth states. This
is different from the situation in the three state model corresponding to
the supersymmetric $t$--$J$ model for which three Bethe Ans\"atze
corresponding to the various possibilities of ordering of the basis
(\ref{bas3})---namely FFB, FBF and BFF---can be constructed \cite{esko:92}.

The fact that there exists a family of four dimensional representations for
this superalgebra allows to introduce a new type of inhomogenous four-state
vertex models by allowing the parameter $b$ to take different values in
different quantum spaces. Studying the Hamiltonian limit of this class of
inhomogenous vertex models leads to systems of electrons with correlated
hopping with a spatially varying parameter. It should be noted however,
that the $R$ matrix $R^{b_1b_2}(\mu)$ becomes proportional to a (graded)
permutation operator for some values of $\mu$ \emph{only} if $b_1=b_2=b$.
The existence of such a shift point is necessary for the construction of a
local Hamiltonian from the transfer matrix. Thus to limit the range of
interaction one should consider a model with a sufficient number of sites
carrying the same representation as the auxiliary space of the monodromy
matrix. A possible example is a single $b'$ ``impurity'' in a chain built
from $R^{bb}$ otherwise. We shall study the effect of such an impurity on
the thermodynamic properties of an correlated electronic system in a
forthcoming paper.

\emph{Note added:} After completion of this work we received a preprint by
P.~B.~Ramos and M.~J.~Martins \cite{rama:96}
who obtain the spectrum (\ref{EWb1b2}) of the
transfer matrix $\tau^{bb}(\mu)$ by applying the algebraic Bethe Ansatz to
the $4\times4$ monodromy matrix $T^{bb}(\mu)$ directly. Their results
concide with ours, it should be noted though that their discussion of
the \emph{unwanted} terms arising in this procedure is not complete.
\section*{Acknowledgment}
%
This work has been supported in part by the Deutsche Forschungsgemeinschaft
under grant no Fr 737/2-1.

\begin{thebibliography}{10}

\bibitem{bglz:95}
A.~J. Bracken, M.~D. Gould, J.~R. Links, and Y.-Z. Zhang,
\newblock \emph{Phys. Rev. Lett.} \textbf{74}, 2768 (1995).

\bibitem{maas:95}
Z.~Maassarani,
\newblock \emph{J. Phys. A} \textbf{28}, 1305 (1995).

\bibitem{esko:92}
F.~H.~L. Essler and V.~E. Korepin,
\newblock \emph{Phys. Rev. B} \textbf{46}, 9147 (1992).

\bibitem{foka:93}
A.~Foerster and M.~Karowski,
\newblock \emph{Nucl. Phys. B} \textbf{396}, 611 (1993).

\bibitem{lai:74}
C.~K. Lai,
\newblock \emph{J. Math. Phys.} \textbf{15}, 1675 (1974).

\bibitem{suth:75}
B.~Sutherland,
\newblock \emph{Phys. Rev. B} \textbf{12}, 3795 (1975).

\bibitem{schl:87}
P.~Schlottmann,
\newblock \emph{Phys. Rev. B} \textbf{36}, 5177 (1987).

\bibitem{karn:94}
I.~N. Karnaukhov,
\newblock \emph{Phys. Rev. Lett.} \textbf{73}, 1130 (1994).

\bibitem{befr:95}
G.~Bed{\"u}rftig and H.~Frahm,
\newblock \emph{J. Phys. A} \textbf{28}, 4453 (1995).

\bibitem{snr:77}
M.~Scheunert, W.~Nahm, and V.~Rittenberg,
\newblock \emph{J. Math. Phys.} \textbf{18}, 155 (1977).

\bibitem{marc:80I}
M.~Marcu,
\newblock \emph{J. Math. Phys.} \textbf{21}, 1277 (1980).

\bibitem{dglz:95}
G.~W. Delius, M.~D. Gould, J.~R. Links, and Y.-Z. Zhang,
\newblock \emph{Int. Jour. Mod. Phys. A} \textbf{10}, 3259 (1995).

\bibitem{rama:96}
P.~B. Ramos and M.~M. J.,
\newblock ``One parameter family of an integrable spl(2|1) vertex model:
  Algebraic bethe ansatz and ground state structure'', preprint
  UFSCARF-TH-96-03, (1996).

\end{thebibliography}
\end{document}